\documentclass[runningheads]{llncs}
\usepackage[T1]{fontenc}
\usepackage{graphicx}
\usepackage{subfigure}
\usepackage{amsmath,amssymb}
\usepackage{booktabs}
\usepackage{dcolumn}
\usepackage{array}
\usepackage{color}
\begin{document}
\title{Preserving Cardiac Integrity: A Topology-Infused Approach to Whole Heart Segmentation}
\titlerunning{Preserving Cardiac Integrity: A Topology-Infused Approach to WHS}
%
\author{Chenyu Zhang\inst{1} \and
Wenxue Guan\inst{4} \and
  Xiaodan Xing\inst{1}\and
 Guang Yang\inst{1,2,3,5}}
 \authorrunning{Chenyu. z et al.}
 \institute{Bioengineering Department and Imperial-X, Imperial College London, London W12 7SL, United Kingdom\\
 \email{\{chenyu.zhang24,x.xing\}@imperial.ac.uk}\and
 National Heart and Lung Institute, Imperial College London, London, United Kingdom\and
 School of Biomedical Engineering \& Imaging Sciences, King's College London, London, United Kingdom
 \\
 \and
 Jilin university,Changchun,China\and
 Cardiovascular Research Centre, Royal Brompton Hospital, London, United Kingdom\\
 \email{gyang@imperial.ac.uk}}
\maketitle
\begin{abstract}
Whole heart segmentation (WHS) supports cardiovascular disease (CVD) diagnosis, disease monitoring, treatment planning, and prognosis. Deep learning has become the most widely used method for WHS applications in recent years. However, segmentation of whole-heart structures faces numerous challenges including heart shape variability during the cardiac cycle, clinical artifacts like motion and poor contrast-to-noise ratio, domain shifts in multi-center data, and the distinct modalities of CT and MRI. To address these limitations and improve segmentation quality, this paper introduces a new topology-preserving module that is integrated into deep neural networks. The implementation achieves anatomically plausible segmentation by using learned topology-preserving fields, which are based entirely on 3D convolution and are therefore very effective for 3D voxel data. We incorporate natural constraints between structures into the end-to-end training and enrich the feature representation of the neural network. The effectiveness of the proposed method is validated on an open-source medical heart dataset, specifically using the WHS++ data. The results demonstrate that the architecture performs exceptionally well, achieving a Dice coefficient of 0.939 during testing. This indicates full topology preservation for individual structures and significantly outperforms other baselines in preserving the overall scene topology. 

\end{abstract}

\section{Introduction}

\begin{figure}[t]
\centering  
\subfigure[]{
\label{Fig.sub.1}
\includegraphics[width=5.2cm,height = 3.8cm]{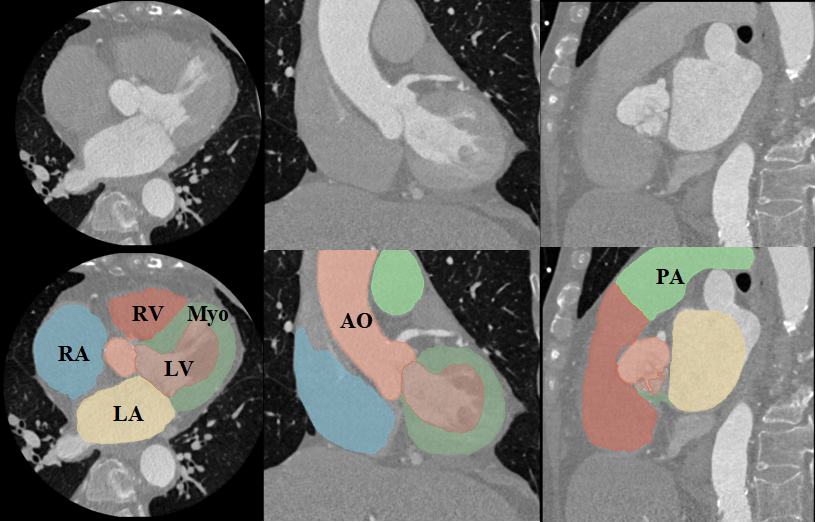}}\subfigure[]{
\label{Fig.sub.2}
\includegraphics[width=5.2cm,height = 3.8cm]{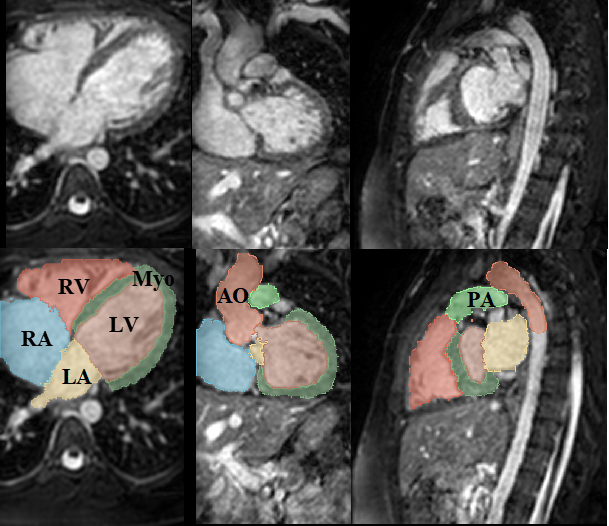}}
\caption{Examples of cardiac images and WHS results: (a) displays the three orthogonal views of a cardiac CT image and its corresponding WHS result, (b) shows example
cardiac MRI data and the WHS result. LV: left ventricle; RV: right ventricle; LA: left atrium; RA: right atrium; Myo: myocardium of LV; AO: ascending aorta; PA: pulmonary
artery.}
\label{fig1}
\end{figure}

In recent decades, significant advances in cardiovascular disease (CVD) research and practice have been made to improve the diagnosis and treatment of heart disease~\cite{faucris.290665703}. Cardiac image segmentation is an important first step in many applications. It partitions the image into several semantically (i.e., anatomically) meaningful regions, based on which quantitative measures can be extracted, such as the myocardial (Myo) mass, wall thickness, left ventricle (LV), right ventricle (RV) volume, left atrium (LA), right atrium (RA), ascending aorta (AO) or the whole aorta, and the pulmonary artery (PA)\cite{Zhuang2019EvaluationOA,Zhuang2019evaluation}, as Fig.~\ref{fig1} shows. Despite its wide range of applications, automated WHS remains challenging, e.g., systole and diastole of the heart, deformation of the heart's shape from pathological and physiological changes, as well as differences in image acquisition equipment and motion artifacts that appear in the clinical data. State-of-the-art deep segmentation methods\cite{Payer2017MultilabelWH,Heinrich2017MRIWH,Wang2017AutomaticWH,Yang20173DCN,Mortazi2017CardiacNETSO,Zhuang2016MSMMA,GAO2023BayeSeg,Zhuang2019MvMM} can automatically learn robust feature representations with satisfactory performance. However, standard deep neural networks cannot learn global structural constraints on semantic labeling, which is often crucial in the biomedical domain. For example, in WHS segmentation, we know in advance that Myo always encloses LV, and another global constraint is mutual exclusion for different labels. 

We propose a new approach to learning these global structural constraints for WHS. In addition, this module is incorporated into training to aid network learning. Our key observation is that a broad class of topological restriction effects, namely containment, and exclusion, can be attributed to certain impermissible combinations of labels of neighboring voxels. Inspired by such observations, we propose a topology-preserving module that encodes constraints into a neural network through a series of convolution operations. Our module is very efficient because of the 3d convolution-based design. Moreover, it can be naturally incorporated into the training of the neural network, e.g., by penalizing constraint-violating voxels with additional losses. The network learns to segment correctly with our module even when a strong baseline, such as nnUNet~\cite{Isensee2020nnUNetAS}, fails. We evaluated the proposed method by conducting experiments on the WHS++ challenge dataset. The results show that our method is efficient, especially on the MRI dataset. It not only strengthens the constraints, but also significantly improves the segmentation quality on the Dice score, Jaccard index, surface-to-surface distance (SD), and Hausdorff Distance (HD) standard metrics.

\section{Related work}
In the era of deep learning, many methods have been proposed to deal with CT and MRI image segmentation of the whole heart. Through~\cite{faucris.290665703}, they can be classified into three categories, which are two-step segmentation, multi-view, and hybrid loss-based models.

\subsubsection{Two-step segmentation}Some deep learning methods rely on a two-step segmentation process in which the region of interest (ROI) is first extracted and then fed into a CNN for subsequent classification\cite{Zreik2016AutomaticSO,Dormer2018HeartCS}. For example, Zreik et al.~\cite{Zreik2016AutomaticSO}proposed a two-step LV segmentation process in which the bounding box of the LV is first detected using the method described in de Vos et al.~\cite{deVos2017ConvNetBasedLO}, and then voxel classification is carried out within the defined bounding box using a patch-based CNN. Recently, FCNs, especially U-net~\cite{Ronneberger2015UNetCN}, have become the method of choice for cardiac CT segmentation. Several of these methods\cite{Payer2017MultilabelWH,Wang2017AutomaticWH,Xu2018CFUNCF,Wang2018AT3}combine a localization network that produces a coarse detection of the heart with a 3D FCN applied to the detected ROIs for segmentation. This allows the segmentation network to focus on anatomically relevant regions and is effective for WHS. Zhuang et al.~\cite{Zhuang2019EvaluationOA} compared a set of methods for WHS that have been evaluated in the MM-WHS challenge\cite{Payer2017MultilabelWH,Heinrich2017MRIWH,Wang2017AutomaticWH,Yang20173DCN,Mortazi2017CardiacNETSO,Galisot2017LocalPA,Xu2018CFUNCF,Wang2018AT3}. The segmentation accuracy of the methods evaluated on the MM-WHS dataset was comparable to that of MRI images, and these methods typically achieve better segmentation accuracy on CT images, mainly due to less variation in image intensity distribution and better image quality between different CT scanners.
\subsubsection{Multi-view}
Another line of research utilizes the volumetric information of the heart by training multi-planar CNNs (axial, sagittal, and coronal views) in a 2D fashion. Examples include Wang et al.~\cite{Wang2017AutomaticWH}and Mortazi et al.~\cite{Mortazi2017CardiacNETSO}where three independent orthogonal CNNs were trained to segment different views. Specifically, Wang et al.~\cite{Wang2017AutomaticWH} additionally incorporated shape context in the framework for the segmentation. In contrast, Mortazi et al.~\cite{Mortazi2017CardiacNETSO}adopted an adaptive fusion strategy to combine multiple outputs utilizing complementary information from different planes.
\subsubsection{Hybrid loss}
Several methods employ a hybrid loss, where different loss functions (such as focal loss, Dice loss, and weighted categorical cross-entropy) are combined to address the class imbalance issue, e.g., the volume size imbalance among different ventricular structures, and to improve the segmentation performance~\cite{Ye2019MultiDepthFN}.

\section{Method}
\begin{figure}
\centering
\includegraphics[width=0.9\textwidth]{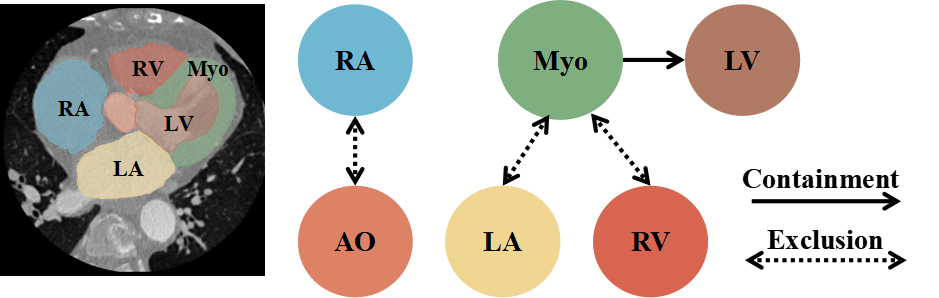}
\caption{ Multi-class topological constraints for WHS} \label{fig3}
\end{figure}

Topological constraints between different foreground classes include two types, containment and exclusion. Referred on the~\cite{Gupta2022LearningTI}, we illustrate these constraints using three class labels,$\alpha$, $\beta$, and $\gamma$. In Fig.~\ref{fig3}, We use a solid arrow from $\beta$ to $\alpha$ to denote the containment relationship. In real applications, e.g., Myo always encloses LV. Classes $\alpha$ and $\gamma$ are mutually exclusive if the voxels of class $\alpha$ and class $\gamma$ cannot be adjacent to each other. We use a dashed double arrow to denote the exclusion relationship. In WHS, there is a clear separation between RA and AO. They are mutually exclusive.

\begin{figure}
\includegraphics[width=\textwidth]{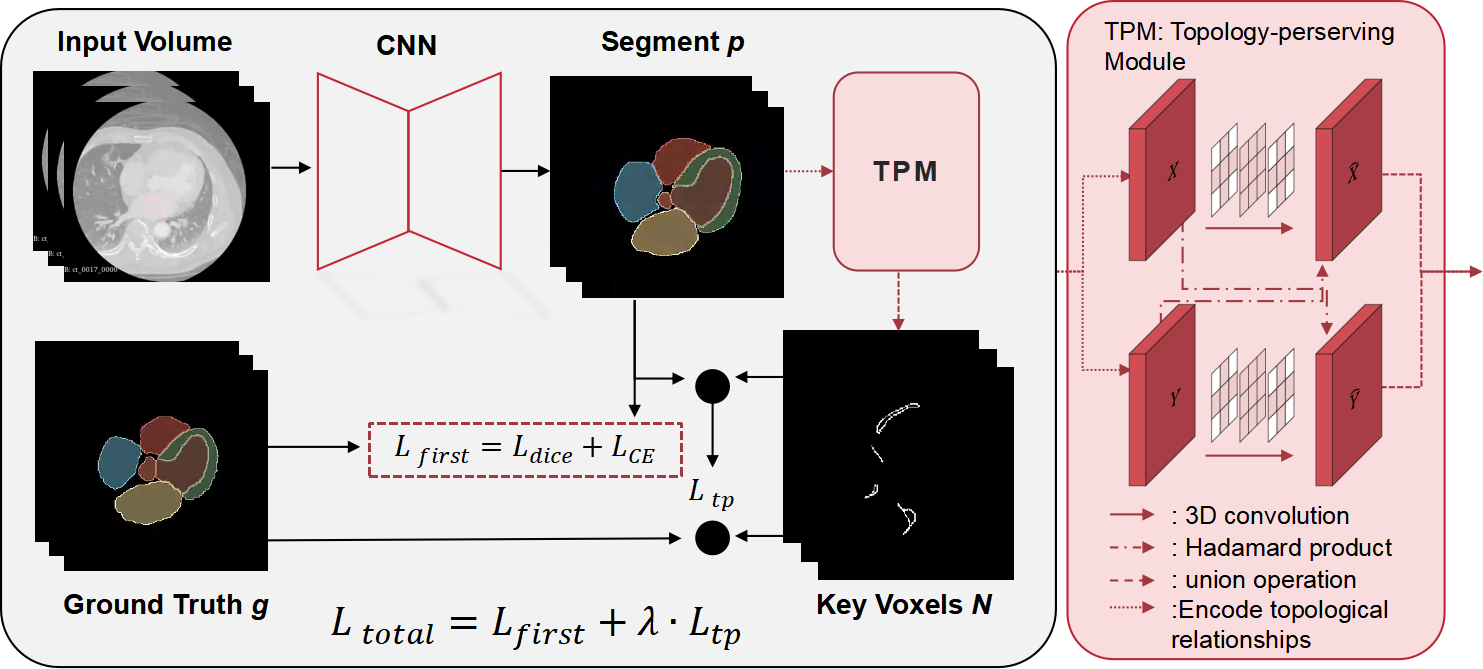}
\caption{ An overview of the proposed method. TPM encodes priori knowledge between the different classes (e.g., Myo and LA classes in the WHS++ dataset follow the exclusion constraint). Key voxels N are identified and used for the new loss.} \label{fig4}
\end{figure}

By proposing a new topology-preserving module, we apply these constraints to CNN training. The idea is to go through all neighboring voxel pairs and identify pairs that violate the desired constraints. We will refer to them as key voxels. We can then merge the module into the training by designing additional losses on these key voxels. Fig.~\ref{fig4} provides an overview of the proposed method.

\subsection{Topology-preserving Module}
After the network predicted the 3-dimensional discrete segmentation, the topology-preserving module encodes the topological relationships defined above. The network resembles that of the UNet~\cite{Ronneberger2015UNetCN}, as it is made up of a contracting encoder branch that is connected through skip connections to the expanding decoder branches. As mentioned before, the key is to forbid certain label combinations from appearing in any pair of adjacent voxels. Our module identifies the pairs that violate the constraints. To encode a priori knowledge as local constraints, this paper creates new labels X and Y to represent exclusion and containment constraints uniformly, and for exclusion constraints the new labels X = $\alpha$ and Y = $\gamma$ and prohibit them from appearing in neighboring voxels. For containment constraints, on the other hand, consider X = $\alpha$ and Y as the connected set of all other labels except $\alpha$ and $\beta$. Then the containment relations $\alpha$ and $\beta$ are equivalent to X = $\alpha$ not touching Y = $\gamma$~\cite{Gupta2022LearningTI}. Afterward, to identify neighboring voxel pairs with label pairs (X, Y) or (Y, X), we performed 3D convolutions on different categories, where the kernel was refined.
In practice, masks can be efficiently expanded using the expansion morphological operation~\cite{Haralick1987ImageAU}. In the expansion process, we convolve a given binary mask with the kernel, as shown in Fig.~\ref{fig3}. We want all voxels in contact with class X or Y to be activated. Afterward, we can find which voxels of X and Y belong to each other's neighborhoods by performing  Hadamard product calculations and union operations on them respectively, which is key voxels N.

\subsection{End-to-End Training}
To incorporate the proposed topology-preserving module into end-to-end training, we propose a method with topology-constrained loss that corrects violations by penalizing key voxels.

Let $p\in{R^{7\times H\times W}}$ be the multi-class likelihood map predicted by the network, where H and W denote the number of height, and width of the image, respectively. $g\in{R^{H\times W}}$ is the ground truth segmentation map with discrete labels that represent Myo, LV, RV, LA, RA, AO, and PA respectively. Among them, We use cross-entropy to calculate and use the mask N obtained from Sec. 3.1, to define ${L_{tp}}$, denoting the additional topology-constrained, as:
\begin{equation}
{L_{tp}} = {L_{CE}}\left( {p \odot N,g \odot N} \right),
\end{equation}
 where $\odot$ denotes the Hadamard product. ${L_{tp}}$ can essentially encode the topological constraints, correct the topological errors, and eventually produce a topologically correct segmentation. The final loss of our method, ${L_{total}}$, is given by:
 \begin{equation}
{L_{total}} = {L_{CE}} + {L_{dice}} + {\lambda {L_{tp}}},
\end{equation}
where ${L_{CE}}$ and ${L_{dice}}$ denote the cross-entropy and dice loss. The loss is controlled by the weights $\lambda$.

\section{Experiments}
\subsection{Dataset}
We evaluated the method on the datasets of the WHS++ challenge. The organizers provided 104 CT and 106 MRI volumes with corresponding manual segmentations of seven whole heart substructures. The volumes were acquired in clinics with different scanners, resulting in varying image quality, resolution, and voxel spacing.
\subsection{Evaluation metrics}
We employed four widely used metrics to evaluate the accuracy of a segmentation result~\cite{Zhuang2013ChallengesAM}: the Dice score, Jaccard index, surface-to-surface distance (SD), and Hausdorff Distance (HD). For WHS evaluation, the generalized metrics were used, which are expected to be more objective\cite{Zhuang2013ChallengesAM,Crum2006GeneralizedOM}.
\subsection{Implementation Details}
We train and test the networks with pytorch~\cite{Paszke2019PyTorchAI} where we perform data augmentations using ITK, i.e., random rotations, random scaling, random elastic deformations, gamma correction augmentation, and mirroring.

We use the Adam optimizer with an initial learning rate of $3\times {10^{ - 4}}$ for all experiments. We define an epoch as the iteration over 250 training batches. The value of $\lambda$ is ${10^{ - 6}}$. Due to memory restrictions coming from the volumetric inputs and the use of 3D convolution kernels, we choose a mini-batch size of 2. All experiments were performed on an Intel Core i9-8950HK-based workstation with a 24GB NVidia Geforce RTX 4090.

\begin{figure}[t]
\centering  
\subfigure[Input]{\rotatebox{90}{\scriptsize{~~~~~~~MRI~~~~~~~~~~~~~~~CT}}
\label{Fig5.sub.1}
\includegraphics[width=0.15\textwidth]{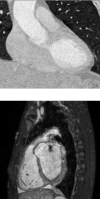}}\subfigure[GT]{
\label{Fig5.sub.2}
\includegraphics[width=0.15\textwidth]{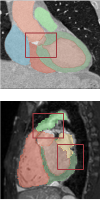}}
\subfigure[UNet]{
\label{Fig5.sub.3}
\includegraphics[width=0.15\textwidth]{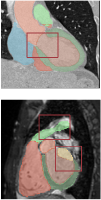}}
\subfigure[3dUNet]{
\label{Fig5.sub.4}
\includegraphics[width=0.15\textwidth]{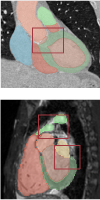}}
\subfigure[nnUNet]{
\label{Fig5.sub.5}
\includegraphics[width=0.15\textwidth]{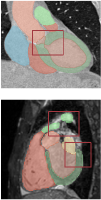}}
\subfigure[ours]{
\label{Fig5.sub.6}
\includegraphics[width=0.15\textwidth]{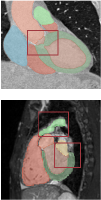}}
\caption{Qualitative results compared with the baselines. The first row is CT data and the second row is MRI data. Colors for the classes correspond to the ones used in Fig.~\ref{fig3}}
\label{fig6}
\end{figure}

\subsection{Baselines and main results}
The comparison baselines consist of the nnUNet~\cite{Isensee2020nnUNetAS}, UNet~\cite{Ronneberger2015UNetCN}, and 3D-UNet~\cite{iek20163DUL}. We use the publicly available codes for these. 
Tab.1 presents the quantitative results of the evaluated algorithms on the CT and MRI datasets. In general, we observe that learning the topological constraint leads to better feature representation and thus better qualitative and quantitative segmentations. Although anatomically the ventricle follows the topological constraint, the available ground truth does not Strictly adhere to it, especially for MR data, manually labeled pixels can be large. So our method did not achieve the best from all metrics. For MRI data, our method also separates it from other categories to a large extent with nnUNet the difference in quantitative metrics is small.
In Fig.~\ref{fig6}, we can see that the network trained using this method has considerably fewer topology violations compared to other baseline networks. Our results show that our proposed method significantly improves the quality of segmentation without any additional post-processing. In the last columns of Fig.~\ref{fig6}, we can see that the proposed method fixes the topological error by enforcing that the LV always surrounds Myo. It can correctly separate these two classes, whereas other methods are not able to do so. Moreover, for the MRI dataset, our method segmented the continuous parts of the PA class. This is also reflected in the quantitative metrics.

\begin{table}[htbp]
    \belowrulesep=0pt
    \aboverulesep=0pt
    \renewcommand{\arraystretch}{1.5}
  \centering
  \caption{Quantitative comparison for CT and MRI dataset.}
    \begin{tabular}{c|c|cccc}
    \toprule[1pt]
    Dataset&Method & Dice  & Jaccard & SD(mm) & HD(mm) \\
    \hline
    \hline
    &UNet   & 0.915 ± 0.086 & 0.852 ± 0.037 & 1.737 ± 0.250 & 25.282 ± 10.813 \\
    CT&3D-UNet   & 0.927 ± 0.032 & 0.867 ± 0.048 & 1.387 ± 0.516 & 31.146 ± 13.203 \\
    &nnUNet & 0.928 ± 0.024 & 0.873 ± 0.042 & \textbf{1.135 ± 0.016} & 4.758 ± 2.045 \\
    &ours  & \textbf{0.949 ± 0.035} & \textbf{0.904 ± 0.046} & 1.139 ± 0.010 & \textbf{4.304 ± 1.978} \\
    \midrule[1pt]
    &UNet   & 0.853 ± 0.043 & 0.762 ± 0.064 & 2.190 ± 0.781 & 30.227 ± 14.046 \\
    MRI&3D-UNet   & 0.893 ± 0.069  & 0.816 ± 0.094  & 1.963 ± 1.012  & 30.201 ± 13.216 \\
    &nnUNet & \textbf{0.902 ± 0.083} & \textbf{0.825 ± 0.046} & 1.723 ± 0.347 & 24.074 ± 11.784 \\
    &ours  & 0.897 ± 0.023 & 0.821 ± 0.059 & \textbf{1.513 ± 0.491} & \textbf{10.001 ± 4.671} \\
    \bottomrule[1pt]
    \end{tabular}%
  \label{tab:addlabel}%
\end{table}%

\begin{figure}
\includegraphics[width=\textwidth]{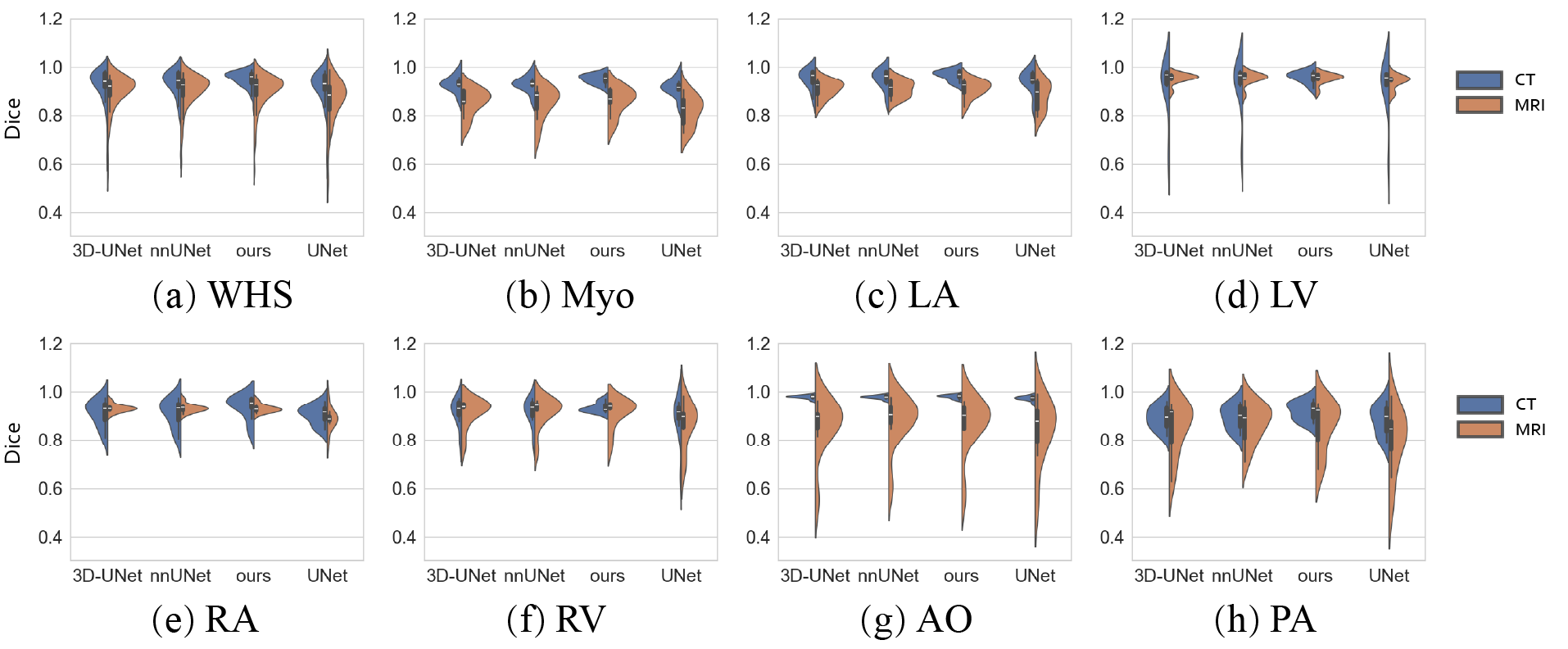}
\caption{Violin plot of Dice scores of the whole heart segmentation on CT and MRI dataset by the four methods.} \label{fig8}
\end{figure}

Fig.~\ref{fig8} shows the violin plots of the evaluated algorithms on CT and MRI data. We can see that they achieve relatively accurate segmentation of all sub-structures of the heart, and the small variation in data density of our method on the Dice score reflects the robustness of the method. By imposing this constraint, our method can better reconstruct both structures, significantly improving the segmentation quality; our method brings the greatest improvement to the left and right ventricles and PA classes, considering that the PA is significantly more variable in shape and appearance.

We can see in Fig.~\ref{fig8} that the segmentation of PA in CT WHS is much worse than the other substructures; in MRI WHS, the segmentation of Myo, AO, and PA seems to be more difficult. This may be because these regions vary more in shape and image appearance from scan to scan. In particular, different pathologies may lead to inhomogeneous intensity of myocardium and blood. Another reason could be the ambiguity of manual descriptions used as ground truth for training learning-based algorithms. This may be greater for MRI data than for CT, as image quality is typically lower (poorer contrast and signal-to-noise ratio) in whole-heart MRI. MRI WHS is typically more challenging than CT WHS and similar conclusions can be drawn for individual substructures and the whole heart when comparing the plots in Fig.~\ref{fig8}.

\section{Discussion and Conclusion}
We introduce a new 3D convolution-based WHS image segmentation module focusing on topological constraints. The module consists of an efficient algorithm to identify topological errors caused by key voxels. We also introduce an additional topology-constrained loss function. By merging this module and the loss function into the training of CNN, we force the network to learn a better representation of the features, thus improving the segmentation quality. The results show that the method can be applied to 3D images and across modalities such as MRI and CT.
\section{Acknowledgement}
Guang Yang was supported in part by the ERC IMI (101005122), the H2020 (952172), the MRC (MC/PC/21013), the Royal Society (IEC\textbackslash NSFC\textbackslash 211235), the NVIDIA Academic Hardware Grant Program, the SABER project supported by Boehringer Ingelheim Ltd, NIHR Imperial Biomedical Research Centre (RDA01), Wellcome Leap Dynamic Resilience, UKRI guarantee funding for Horizon Europe MSCA Postdoctoral Fellowships (EP/Z002206/1), and the UKRI Future Leaders Fellowship (MR/V023799/1).
%
%
%
\bibliographystyle{splncs04}
\bibliography{mybibliography}
%




\end{document}